\documentclass{article}
\usepackage{spconf,graphicx,booktabs,amssymb,multirow}
\usepackage[hidelinks]{hyperref}

\newcommand{\demourl}{https://pnlong.github.io/SPDMX/}

\title{SPDMX: Large-Scale Open Stem Audio from Symbolic Music}

\name{Phillip Long, Eduardo Escoto, Julian McAuley, Zachary Novack}
\address{University of California, San Diego Computer Science and Engineering Department}

\begin{document}
\maketitle

\begin{abstract}
Large public datasets have driven recent progress in machine learning, from text and vision to speech and general audio.
Music remains an outlier, as copyright blocks redistribution of training audio, in particular for modern tasks that need \emph{individual stems} rather than mixtures.
Existing multitrack resources trade off realism, scale, and licensing, as synthetic alternatives rely on proprietary VSTs or naive open rendering that yields weak stems.
We introduce \textbf{SPDMX}, an openly-licensed stem corpus rendered from the public-domain PDMX dataset.
SPDMX comprises $5{,}469$ song-hours ($2{,}734$ song-hours, or $34.8\%$, of which are multitrack scores) at $44.1$\,kHz under CC~BY~4.0, with paired MIDI and identifiers that link our dataset to PDMX's original metadata.
To our knowledge, SPDMX's multitrack subset constitutes the largest openly-licensed synthetic multitrack subset by song-hours.
We then demo SPDMX as a dataset for both source separation and transcription, showing how its increased scale can enable better training of models on synthetic data.
We release SPDMX, as well as the open creation pipeline for reuse on other symbolic music datasets.\footnotemark[1]
\end{abstract}
\footnotetext[1]{\url{\demourl}}


\vspace{-1em}

\section{Introduction}
\label{sec:intro}

Massive public datasets have driven much of the progress in machine learning over the past decade, from text~\cite{raffel2020exploring,gao2020pile} and images~\cite{deng2009imagenet,schuhmann2022laion} to speech and general audio~\cite{gemmeke2017audio,panayotov2015librispeech}.
Music has still struggled to follow the same path, as copyright law blocks the legal redistribution of recordings that would form such corpora.
In particular, this has been notably apparent for tasks that need \emph{individual stems}, such as source separation, transcription, or accompaniment generation, which have been relegated to multi-year-old datasets in this absence.
Legally redistributable multitrack data remains scarce.

Existing resources trade off realism, scale, and redistributability (Table~\ref{tab:datasets}). 
Recorded datasets such as MUSDB18~\cite{rafii2019musdb18}, MedleyDB~\cite{bittner2014medleydb,bittner2016medleydb}, and MoisesDB~\cite{pereira2023moisesdb} are realistic but small or have restricted licenses, while the larger Slakh2100~\cite{manilow2019cutting} still uses non-redistributable commercial instruments on non-licensed MIDI.
PDMX~\cite{long2024pdmx,xu2024generating} offers greater scale, but naive General MIDI (GM) rendering yields suboptimal realism.
No prior resource jointly offers scale, license openness, and passable realism.

\begin{table}[t]
\centering
\setlength{\tabcolsep}{3.5pt}
\caption{Multitrack resources ($N$\,=\,songs, $h$\,=\,mixture hours; T: R\,=\,recorded, S\,=\,synthetic; D\,=\,openly-licensed download; P\,=\,synthesis without proprietary instruments). 
SPDMX includes single-stem songs.}
\label{tab:datasets}
\small
\begin{tabular}{@{}lrrccc@{}}
\toprule
\textbf{Dataset} & $N$ & $h$ & T & D & P \\
\midrule
MUSDB18~\cite{rafii2019musdb18} & 150 & 10 & R & $\times$ & --- \\
MedleyDB$^{\ddagger}$~\cite{bittner2014medleydb,bittner2016medleydb} & 196 & 13.8 & R & $\times$ & --- \\
MoisesDB~\cite{pereira2023moisesdb} & 240 & 14.4 & R & $\times$ & --- \\
Slakh2100$^{\dagger}$~\cite{manilow2019cutting} & 2{,}100 & 145 & S & \checkmark & $\times$ \\
CocoChorales~\cite{wu2022chamber} & 240{,}000 & 1{,}411 & S & \checkmark & \checkmark \\
EnsembleSet~\cite{sarkar2022ensembleset} & 80 & 6.15 & S & \checkmark & $\times$ \\
\midrule
\textbf{SPDMX (All)} & 253{,}982 & 5{,}469 & S & \checkmark & \checkmark \\
\textbf{SPDMX (Multitrack)} & 88{,}403 & 2{,}734 & S & \checkmark & \checkmark \\
\bottomrule
\end{tabular}\\[0.4em]
\footnotesize{$^{\dagger}$Suspect CC~BY audio, but proprietary VST renderers.\\
$^{\ddagger}$MedleyDB $\cup$ MedleyDB 2.0.}
\end{table}

To address this gap, we propose \textbf{SPDMX}, a rendering of PDMX comprising $253{,}982$ songs, $490{,}159$ mono stems, and $5{,}469$ song-hours at $44.1$\,kHz under CC~BY~4.0.
Its $88{,}403$ multitrack songs total $2{,}734$ song-hours, exceeding Slakh2100 by over an order of magnitude.
We clean each song's MIDI and select category-specific FluidSynth or MIDI-DDSP~\cite{wu2022midi} recipes, uncovered through a small listening pilot, to produce a reusable open pipeline.

We then validate SPDMX as a training and evaluation dataset for two common multitrack MIR tasks, source separation and transcription.
At fixed-budgets, SPDMX presents a viable alternative or supplementary data source for synthetic training, generally performing at or beyond training purely on Slakh2100.
We hope SPDMX can support a wide range of multitrack tasks, including separation, stem-conditioned generation, MIDI-to-audio synthesis, transcription, and joint symbolic-audio research without raising copyright concerns.
The dataset, samples, and code are on our project page.\footnotemark[1]

\section{Related work}
\label{sec:prior}

\noindent\textbf{Recorded multitrack corpora.}
MUSDB18~\cite{rafii2019musdb18}, MedleyDB~\cite{bittner2014medleydb,bittner2016medleydb}, and MoisesDB~\cite{pereira2023moisesdb} provide high-quality studio stems but at impractical scales for training, while chamber sets such as URMP~\cite{li2018creating} offer realistic ensembles at still smaller scale.
All options remain limited in hours and/or access.
SPDMX instead prioritizes open scale, with thousands of song-hours of redistributable stems with aligned scores, achieved through synthetic generation aligned with human perception.

\noindent\textbf{Synthetic multitrack datasets.}
Slakh2100~\cite{manilow2019cutting} set the standard for this synthesis-based approach, rendering Lakh MIDI~\cite{raffel2016learning} into $\sim$145 song-hours of multitrack audio, but depends on proprietary sample libraries that cannot be redistributed.
EnsembleSet~\cite{sarkar2022ensembleset} and SynthSOD~\cite{garcia2025synthsod} similarly target chamber or orchestral stems with commercial libraries.
As an alternative, neural synthesizers such as MIDI-DDSP~\cite{wu2022midi} (trained on URMP~\cite{li2018creating}) improve realism for selected monophonic instruments and underlie generative releases like CocoChorales~\cite{wu2022chamber}. 
SPDMX instead renders public-domain PDMX repertoire with an open FluidSynth/MIDI-DDSP synthesis pipeline, delivering stems, mixtures, and aligned scores at a scale previous resources do not provide.
Among the resources surveyed in Table~\ref{tab:datasets}, its $2{,}734$-hour multitrack subset is the largest openly-licensed synthetic collection by song-hours.

\section{SPDMX Dataset}
\label{sec:dataset}

\begin{figure}[t]
\centering
\includegraphics[width=\linewidth]{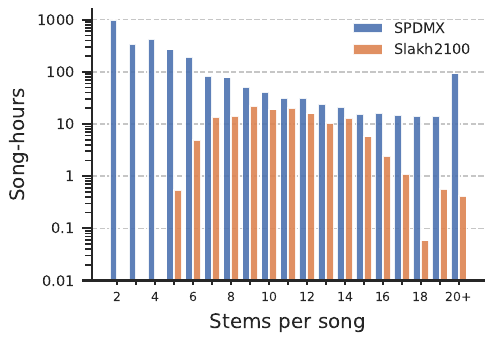}
\caption{Song-hours versus stems per song for SPDMX and Slakh2100 (log $y$).
SPDMX singletrack songs are omitted; Slakh2100 includes all $2{,}100$ tracks ($145$\,h).
Multitrack SPDMX totals $2{,}734$ song-hours across $88{,}403$ songs.}
\label{fig:stems_per_song}
\end{figure}

SPDMX comprises $253{,}982$ songs ($5{,}469$ song-hours) as $490{,}159$ mono FLAC stems paired with MIDI. 
Shared song identifiers enable PDMX metadata retrieval.
We use ``song-hours'' to denote the total duration only looking at mixes, to not overcount songs with many stems.
Of these, $88{,}403$ songs ($34.8\%$; $2{,}734$ song-hours, $50.0\%$ of hours) have $\geq 2$ stems. 
Fig.~\ref{fig:stems_per_song} compares multitrack song-hours by stem count against Slakh2100: even excluding SPDMX's $165{,}579$ single-stem songs, the multitrack subset alone already exceeds Slakh2100's full $145$\,h inventory.
We additionally identify a 632-song BDGP subset (bass, drums, guitar, and piano present) to support matched comparisons against Slakh2100.
As we will discuss in subsequent sections, all rendering uses only open synthesizers, with audio released under CC~BY~4.0.

We describe SPDMX's construction in three stages: first we clean PDMX scores, then select a renderer family (FluidSynth or MIDI-DDSP) for each instrument category, and finally compile the full dataset from those stems.

\subsection{Initial data cleaning}
\label{ssec:cleaning}

\begin{figure}[t]
\centering
\includegraphics[width=\linewidth]{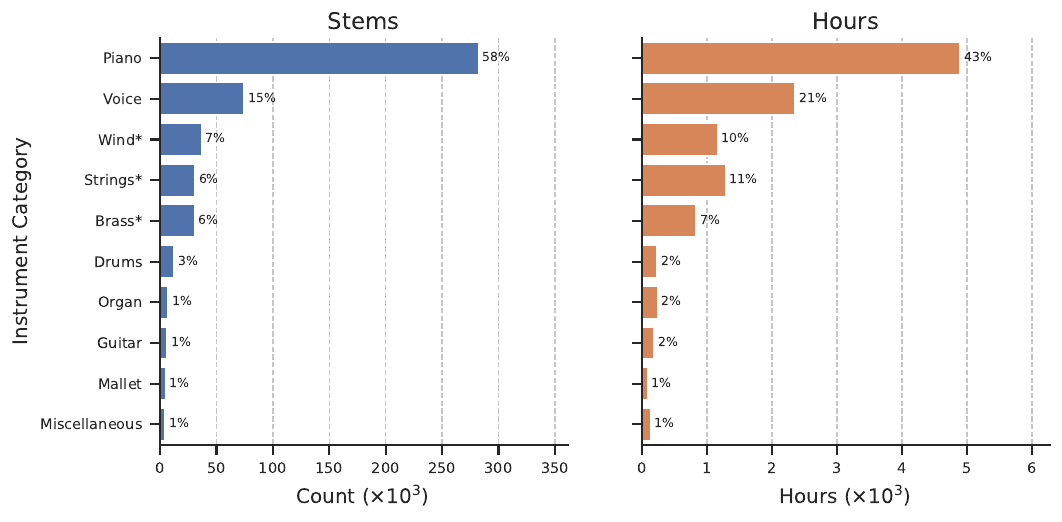}
\caption{SPDMX stem counts and non-silent hours by instrument category.
Categories marked $*$ are MIDI-DDSP eligible in our frozen hybrid recipe (strings, wind, and brass).}
\label{fig:gm_programs}
\end{figure}

Direct use of PDMX data initially posed an issue, as MIDI programs often disagree with track names (e.g., a named violin track may be encoded as piano in the MIDI program).
We thus correct GM programs from track name aliases and further drop tracks with no note events.
Fig.~\ref{fig:gm_programs} shows the corrected inventory by stem count and non-silent hours.
For each rendered stem, we sum $0.25$\,s windows whose RMS amplitude is at least $0.01$ ($\approx -40$\,dBFS), excluding near-silent gaps between and on the tail-end of notes.
Much of SPDMX concentrates in piano, voice, drums, and common orchestral families.

Although this cleaning clarifies the symbolic source of SPDMX, it fails to resolve the acoustic limitations of using a single GM bank to render the whole dataset.
Naive GM rendering remains dry and synthetic, particularly apparent in winds, brass, and strings, where standard soundfonts fail to create realistic acoustic sustain.
This issue motivates our listening test to assess which open-source sound synthesis tools can be used to improve realism.

\subsection{Listening study}
\label{ssec:listening}

We use a small listening pilot to select a render recipe for each of the following instrument categories: drums, organ, guitar, piano, voice, mallet, strings, brass, wind, and miscellaneous.
We synthesize each track as a mono stem at $44.1$\,kHz.
First off, the \emph{basic} soundfont strategy uses one standard GM bank (\texttt{SGM-V2.01}).
The \emph{varied} strategy samples from a shortlist of redistributable soundfonts per category, including \texttt{FluidR3\_GM}, \texttt{GeneralUser GS}, \texttt{Arachno}, and \texttt{Airfont}, and also applies a light reverb FX profile, aiming to replicate Slakh2100's timbral variety without its proprietary libraries.
For eligible monophonic strings, brass, and winds, we also experiment with using MIDI-DDSP~\cite{wu2022midi} to replace soundfonts. 
In such cases, we fall back to FluidSynth for polyphonic parts and instruments outside the MIDI-DDSP (URMP) inventory.
Notably, MIDI-DDSP synthesizes at $16$\,kHz, so we upsample to $44.1$\,kHz so that all stems share one sample rate and easily mix together.
Nonetheless, those stems remain inherently band-limited relative to native $44.1$\,kHz audio.
Voice-class tracks use choir or voice soundfonts, and lyric singing synthesis is out of scope for the present work.

\begin{figure}[t]
\centering
\includegraphics[width=\linewidth]{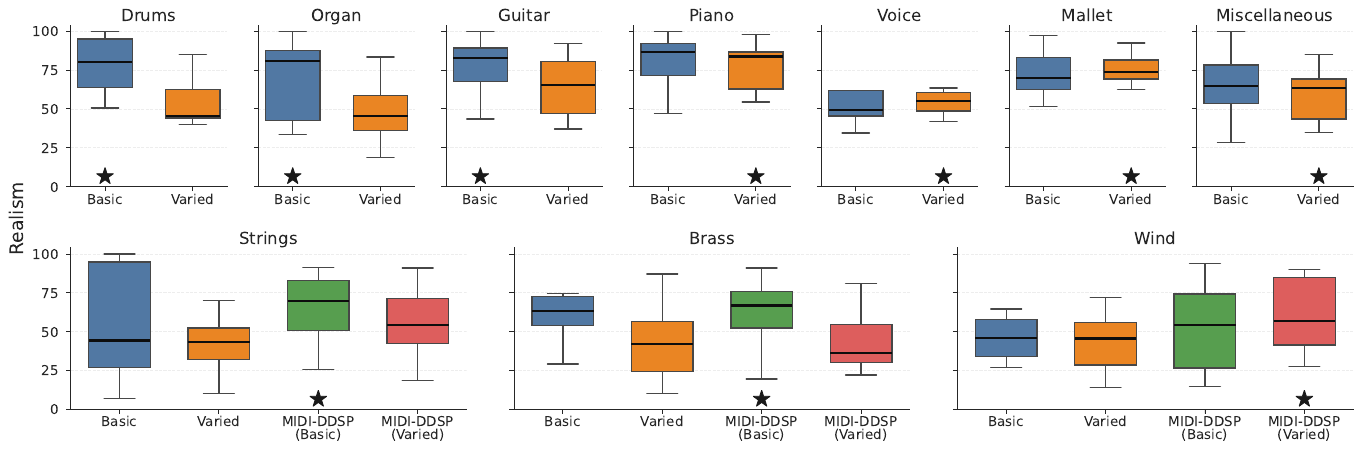}
\caption{Pilot listening ratings of realism ($0$--$100$) by instrument category for basic, varied, and MIDI-DDSP hybrids.
Boxes are per-listener means ($N{=}8$; two stems per category).
MIDI-DDSP panels appear only where neural synthesis can apply.
Stars mark the frozen SPDMX recipe.}
\label{fig:listening}
\end{figure}

We ask listeners to compare four families on perceived realism: basic, varied, MIDI-DDSP with basic fallbacks, and MIDI-DDSP with varied fallbacks.
Each trial scores a $10$\,s stem excerpt for perceived realism ($0$--$100$), and each category has two trials.
$N{=}8$ listeners completed the study (Fig.~\ref{fig:listening}).

We use these ratings as design guidance. 
Specifically, we choose the highest mean by default, but when basic and varied are close, we retain varied to preserve timbral diversity.
SPDMX uses \textit{basic} for drums, organ, and guitar, and \textit{varied} for piano, voice, mallet, and miscellaneous classes; strings and brass use MIDI-DDSP with basic, and wind uses MIDI-DDSP with varied.

\subsection{Final assembly}
\label{ssec:assembly}

For every cleaned song, we route each stem through the appropriate rendering recipe. 
Independent per-stem loudness matching erases the inherent MIDI velocity balance between separate parts.
To address this, we first loudness-normalize each stem toward $-23$\,LUFS~\cite{series2011algorithms} with peak limiting.
Then, we scale each stem by the ratio of that track's maximum note-on velocity to the global maximum velocity.
Finally, we uniformly apply a single anti-clip peak gain on the stem sum to prevent clipping, which ensures that SPDMX is linearly summable (i.e.~that the mix is the pure sum of the stems)~\cite{manilow2019cutting}.

We recognize several limitations in our construction of SPDMX.
About $65\%$ of songs are single-stem, with piano dominating this subset.
Additionally, MIDI comes from engraved scores with quantized timing and dynamics.
All stems are mono, which erases the spatial realism of instruments like drums and piano.
We note our small, synthetic-reference listening study does not establish corpus-wide absolute realism given the sample size.
MIDI-DDSP stems remain band-limited after upsampling from $16$\,kHz.

\section{Experiments}
\label{sec:experiments}

We demonstrate SPDMX's utility through two experiments with fixed budgets: source-separation and transcription.
Here, we seek to understand whether models can perform well on SPDMX in-domain, and how strongly training with SPDMX translates to performance changes across synthetic and recorded domains.

\subsection{Source separation}
\label{ssec:separation}

We train four-source Hybrid Demucs~\cite{defossez2021hybrid} from random initialization on \emph{bass}, \emph{drums}, \emph{guitar}, and \emph{piano} (BDGP) mixtures using the Slakh2100 four-stem protocol.
We keep only songs that contain all four targets.
The 632 SPDMX songs are shuffled at the song level into 600 training and 32 validation songs.
We train three models for $32{,}000$ updates with batch size 8 on 10-second crops with mono duplicated to stereo: Slakh2100 alone, SPDMX alone, and their concatenated union, sampled uniformly by song.
We report SI-SDR~\cite{leroux2019sdr}, which measures how closely an estimated stem matches its reference while ignoring overall loudness.
We evaluate on Slakh2100 and SPDMX mixtures, and on standard real recorded corpora: MUSDB18~\cite{rafii2019musdb18}, MedleyDB~\cite{bittner2014medleydb,bittner2016medleydb}, and MoisesDB~\cite{pereira2023moisesdb}.

\begin{table}[t]
\centering
\setlength{\tabcolsep}{4pt}
\caption{Hybrid Demucs SI-SDR~\cite{leroux2019sdr} ($\uparrow$, dB; single matched-step run).
BD/GP average bass--drums/guitar--piano; MUSDB18 reports BD only.
SPDMX evaluation set uses 32 songs.}
\label{tab:separation}
\small
\begin{tabular}{@{}llccc@{}}
\toprule
\textbf{Evaluation} & & \textbf{Slakh} & \textbf{SPDMX} & \textbf{Union} \\
\midrule
\multirow{2}{*}{Slakh2100}
  & BD & $\mathbf{10.3}$ & $1.4$ & $10.0$ \\
  & GP & $\mathbf{3.7}$ & $-4.7$ & $3.3$ \\
\midrule
\multirow{2}{*}{SPDMX}
  & BD & $0.2$ & $\mathbf{9.5}$ & $7.5$ \\
  & GP & $-4.1$ & $\mathbf{3.9}$ & $3.3$ \\
\midrule
MUSDB18
  & BD & $-1.7$ & $\mathbf{-0.5}$ & $-0.7$ \\
\midrule
\multirow{2}{*}{MoisesDB}
  & BD & $1.2$ & $1.2$ & $\mathbf{2.4}$ \\
  & GP & $\mathbf{-9.8}$ & $-10.2$ & $\mathbf{-9.8}$ \\
\midrule
\multirow{2}{*}{MedleyDB}
  & BD & $-3.5$ & $-4.0$ & $\mathbf{-2.5}$ \\
  & GP & $\mathbf{-11.0}$ & $-11.9$ & $-11.2$ \\
\bottomrule
\end{tabular}
\end{table}

Table~\ref{tab:separation} reports bass--drums (BD) and guitar--piano (GP) means separately.
In general, we find that single dataset training leads to predictably strong in-domain performance with more suspect domain extrapolation, with both the Slakh2100- and SPDMX-trained models showing comparable performance (the SPDMX model is better at MUSDB18 and is comparable on MoisesDB, with the Slakh2100 model better at MedleyDB).
Interestingly, we find that training on the union of both datasets represents the best of both worlds, maintaining similar performance on in-domain data but better generalization on MoisesDB.
Overall, this shows that SPDMX can at least be used as a standard source separation dataset alongside Slakh2100.

\vspace{-0.8em}

\subsection{Transcription}
\label{ssec:transcription}

Because SPDMX pairs mixtures with MIDI, we also ask whether it can be used for multi-instrument automatic music transcription~\cite{gardner2022mt3,chang2024yourmt3}.
We train a YourMT3-style encoder--decoder (T5-small, trained from scratch) for $100{,}000$ steps on mixture audio with aligned MIDI, comparing Slakh2100-redux against SPDMX songs with $\geq 2$ stems.
We evaluate instrument-agnostic onset F1 on non-drum notes (pitch and onset within $\pm 50$\,ms; offsets ignored) and Multi F1, which includes drums and requires pitch, onset, offset, and instrument matches (drum offsets are ignored), on the official Slakh2100 test split and a held-out SPDMX split.
Note that this setup is intentionally small in training budget and omits the pretrained recipes used in published MT3-style systems, so absolute scores remain low.

\begin{table}[t]
\centering
\caption{Fixed-budget multi-instrument transcription results.
Track mean $\pm$ $95$\% CI.}
\label{tab:transcription}
\begin{tabular}{ll|cc}
\toprule
\textbf{Train} & \textbf{Test} & \textbf{Onset F1}~$\uparrow$ & \textbf{Multi F1}~$\uparrow$ \\
\midrule
Slakh2100 & Slakh2100 & $0.072{\pm}0.005$ & $0.083{\pm}0.006$ \\
Slakh2100 & SPDMX & $0.041{\pm}0.001$ & $0.010{\pm}0.001$ \\
SPDMX & Slakh2100 & $0.075{\pm}0.006$ & $0.066{\pm}0.006$ \\
SPDMX & SPDMX & $\mathbf{0.265{\pm}0.006}$ & $\mathbf{0.191{\pm}0.006}$ \\
\bottomrule
\end{tabular}
\end{table}

In Tab.~\ref{tab:transcription}, we see that on the Slakh2100 test split, Slakh- and SPDMX-trained models reach onset F1 near $0.07$ under equal training budgets.
However on the SPDMX test set, Slakh-trained models are much worse
($0.041$ onset / $0.010$ multi) than the in-domain model ($0.265$ onset / $0.191$ multi), indicating that Slakh training may show lower generalization performance than with SPDMX.

\section{Conclusion}
\label{sec:conclusion}

SPDMX brings public-domain PDMX scores into the audio domain as a large CC~BY~4.0 stem corpus: $253{,}982$ songs, $490{,}159$ mono stems, and $5{,}469$ song-hours at $44.1$\,kHz, with paired MIDI and identifiers that link our dataset to PDMX's original metadata.
$88{,}403$ songs ($34.8\%$; $2{,}734$ song-hours) include multitrack data, well beyond Slakh2100's full inventory and resulting in the largest public domain multitrack dataset to date.
Our listening-guided, open FluidSynth/MIDI-DDSP pipeline makes the release reproducible and extensible to other symbolic datasets. 
Fixed-budget experiments show that SPDMX can be used as both training and evaluation data for source separation and transcription tasks.
We hope SPDMX will support a wide breadth of multitrack and paired audio-MIDI tasks, all without compromising licensing concerns and permitting full open release of such research endeavors.

\section{Compliance with Ethical Standards}
\label{sec:ethics}

SPDMX is derived from public-domain PDMX scores; the rendered audio, MIDI, and release tables are distributed under CC~BY~4.0.
Rendering uses soundfonts selected for redistribution-compatible terms and open MIDI-DDSP weights (Apache~2.0; trained on URMP), not closed commercial VST sample libraries.
The listening study involved adult volunteers rating synthetic audio excerpts.

\bibliographystyle{IEEEbib}
\makeatletter
\let\jazzsamba@thebibliography\thebibliography
\renewcommand{\thebibliography}[1]{%
  \jazzsamba@thebibliography{#1}%
  \setlength{\itemsep}{0pt plus 0.05pt}%
  \setlength{\parsep}{0pt}%
}
\makeatother
\bibliography{main}

\end{document}